\documentclass{article}
\usepackage{INTERSPEECH2019,amsmath,graphicx}
\usepackage[none]{hyphenat}
\usepackage{enumitem}
\usepackage{booktabs}
\usepackage{seqsplit}

\newcommand{\dq}[1]{``#1''}

\graphicspath{ {figures/} }
\usepackage{siunitx}

\usepackage{xcolor}

\title{AUTOMATIC PREDICTION OF SUICIDAL RISK IN MILITARY COUPLES USING MULTIMODAL INTERACTION CUES FROM COUPLES CONVERSATIONS}

\name{Sandeep Nallan Chakravarthula$^{\star}$ \quad Md Nasir$^{\star}$ \quad Shao-Yen Tseng$^{\star}$ \quad Haoqi Li$^{\star}$ \quad Tae Jin Park$^{\star}$ \\ \qquad Brian Baucom$^{\dagger}$ \qquad Craig J. Bryan$^{\dagger\ddag}$ \qquad Shrikanth Narayanan$^{\star}$ \qquad Panayiotis Georgiou$^{\star}$}
\address{$^{\star}$ Dept. of Electrical and Computer Engineering, University of Southern California 
\\ $^{\dagger}$ Dept. of Psychology, University of Utah
\\ $^{\ddag}$ National Center for Veterans Studies, Utah}

\begin{document}\sloppy
\ninept

\maketitle

\begin{abstract}
Suicide is a major societal challenge globally, with a wide range of risk factors, from individual health, psychological and behavioral elements to socio-economic aspects.
Military personnel, in particular, are at especially high risk.
Crisis resources, while helpful, are often constrained by access to clinical visits or therapist availability, especially when needed in a timely manner.
There have hence been efforts on identifying whether communication patterns between couples at home can provide preliminary information about potential suicidal behaviors, prior to intervention.
In this work, we investigate whether acoustic, lexical, behavior and turn-taking cues from military couples' conversations can provide meaningful markers of suicidal risk.
We test their effectiveness in real-world noisy conditions by extracting these cues through an automatic diarization and speech recognition front-end.
Evaluation is performed by classifying 3 degrees of suicidal risk: \textit{none}, \textit{ideation}, \textit{attempt}.
Our automatic system performs significantly better than chance in all classification scenarios and we find that behavior and turn-taking cues are the most informative ones.
We also observe that conditioning on factors such as speaker gender and topic of discussion tends to improve classification performance.
\end{abstract}
\begin{keywords}
Suicidal Risk, Couples Conversations, Prosody, Behavior, ASR
\end{keywords}
\vspace{-0.125cm}
\section{Introduction}
\label{sec:intro}
%\vspace{-0.125cm}
According to the Centers for Disease Control and Prevention, 39,518 suicides were reported in 2011, making suicide the 10th leading cause of death overall and the leading non-natural cause of death for Americans\footnote{https://www.cdc.gov/nchs/fastats/suicide.htm}.
Among military personnel, rates of suicide are even higher \cite{kaplan2007suicide}.
One of the biggest challenges to improving the success of suicide prevention efforts in the military is the absence of reliable methods for predicting who will engage in suicidal behaviors and when they will do so.
This restricts our ability to identify at-risk military personnel and ensure that they are getting the best available and most appropriate treatment for their psychological symptoms.
It also impacts not only them but also their spouses who are at increased risk for a wide range of psychological and physical health symptoms \cite{park2011military,sayers2011family}.

Clinical interviews and surveys are the best available methods for identifying if and when a person is at increased or heightened risk.
However, these methods do not work well for measuring suicide risk in soldiers\footnote{https://theactionalliance.org/sites/default/files/agenda.pdf}.
The major limitations of clinical interviews are that they require in-person interaction with a health care professional and that most service members who die by suicide do not interact with such professionals immediately prior to the event.
Likewise, surveys about suicidal thoughts and feelings are not informative if military personnel are unwilling to acknowledge or are unaware of their psychological distress.
The ability to assess risk directly at home is, therefore, considered valuable, for which machine learning (ML) is being examined as a viable platform.

There is significant literature on using ML for identifying attributes related to suicidal risk; we refer readers to Burke et al. \cite{burke2018use} for a comprehensive review.
Most works either deal with static, non-interactive scenarios such as microblog posts \cite{braithwaite2016validating,cheng2017assessing} and written answers \cite{cook2016novel,pestian2016controlled}, or with interactive but highly structured settings such as interviews with therapists \cite{france2000acoustical} and social workers \cite{scherer2013investigating,venek2017adolescent}.
Existing works typically use information from only one modality such as text \cite{cook2016novel,pestian2016controlled} or audio \cite{france2000acoustical,scherer2013investigating,gideon2019emotion}.
Recently, there have been efforts on using multimodal approaches for quantifying suicidal risk \cite{venek2017adolescent,pestian2017machine}, which is the topic of interest in our work as well.
However, despite this progress, these approaches are often constrained by heavy manual supervision that ensures clean, directly usable features but also greatly limits the scope of their deployment in real-world conditions.
As a result, it is not feasible to deploy these methods at home for suicidal risk assessment.

Distressed couples conversations, which have been well studied within the broad realm of family studies \cite{heavey1993gender,christensen2004,baucom2010gender}, offer a potential setting for performing such assessments.
They have been well analyzed using ML-based computational approaches that have been found to be useful across a variety of behavioral and health domains \cite{narayanan2013behavioral,bone2017signal}. 
For instance, in Couples Therapy \cite{christensen2004}, multiple works have effectively quantified behaviors related to speakers' mental states such as \textit{Blame}, \textit{Positive} and \textit{Sadness} using the speaker's language \cite{georgiou2011s} and vocal traits \cite{black2013toward}.
Similarly, in Cancer Care \cite{reblin2018everyday} interactions, lexical and acoustic cues have been found to be useful in predicting \textit{Hostile} and \textit{Positive} behaviors \cite{chakravarthula2019predicting}.

In this paper, we investigate whether military personnel's conversations with their spouses at home can provide useful markers of their suicidal risk.
We compute various multimodal features relating to behavior, emotion and turn-taking in order to build a comprehensive profile of their communication patterns.
Finally, we test the effectiveness of our work in real-world conditions by extracting all of our features from raw, noisy data in an ecologically-meaningful manner.
Our system uses an automatic diarization and speech recognition front-end with operating conditions that require limited manual supervision and is, hence, readily deployable.
%\vspace{-0.5cm}

\section{Dataset}
\label{sec:data}
%\vspace{-0.125cm}
Our dataset consists of 62 mixed-sex couples, a total of 124 individuals.
They were recruited for a study of behavioral and cognitive markers of suicide risk among geographically dispersed military service members.
The study criteria required that they be in the National Guard, a Reserve Component, or a recent Veteran who served during the Operation Enduring Freedom / Operation Iraqi Freedom era, be married/cohabitating, at least 18 years old, be fluent in English and have reliable internet access at home.

Based on their history of suicidal behaviors, each person was assigned one of 3 labels: (1) \textit{none} if they had no history, (2) \textit{ideation} if they had experienced suicide thoughts but did not act on it and (3) \textit{attempt} if they had attempted suicide in the past.
According to the World Health Organization, a prior attempt is the most important risk factor for suicide\footnote{https://www.who.int/en/news-room/fact-sheets/detail/suicide}; hence, these labels represent the degree of suicidal risk, from \textit{none} representing no risk to \textit{attempt} representing severe risk.
Table \ref{tab:data_labels} shows the demographics of the participants.

\begin{table}[h]
\centering
\begin{tabular}{cccc}
\toprule
\textbf{Gender \textbackslash Label} & \textit{none} & \textit{ideation} & \textit{attempt} \\ \midrule
Husband & 31 & 22 & 9 \\
Wife & 34 & 15 & 13 \\ \bottomrule
\end{tabular}
\caption{Suicidal risk demographics of 62 couples in our data}
\label{tab:data_labels}
\end{table}

As part of their participation, couples completed 2 relationship-change (RC) and 1 reasons-for-living (\textit{RFL}) conversations or \dq{sessions}, in their homes, each one video-recorded for 10 minutes.
In the \textit{RFL} session, they were asked to discuss what they found meaningful, or what their reasons for living were.
In the RC sessions, they were asked to discuss one of their top areas of discontent and conflict in their relationship, with each person getting to select the topic in one session.
We denote the session where the wife picked the topic as \textit{W-Conflict} and the one where the husband picked as \textit{H-Conflict}.
We obtained audio streams from all but one session of one couple, resulting in 370 sessions.

In general, the audio quality was observed to be good; however, some couples recorded their interactions in a noisy environment or did not sit close enough to the recorder to be clearly intelligible.
Nevertheless, we retained all their samples consistent with the goal of this work of using data reflecting real-world operational use conditions.
\vspace{-0.5cm}

\section{Feature Extraction}
\label{sec:features}

\subsection{Automatic Data Processing}
\label{ssec:processing}

\subsubsection{Diarization}
\label{sssec:diar}
% The first, important step towards transcribing couples conversations is speaker diarization, i.e. classifying speaker identities to figure out \dq{who spoke when} during the interaction.
The first, important step towards automatically analyzing couples conversations is speaker diarization, i.e. identifying \dq{who spoke when} during the interaction.
We employ the x-vector \cite{snyder2018x} based diarization system proposed in \cite{park2019second} for extracting speaker embeddings in each speech segment and clustering them using the spectral clustering approach described in \cite{park2019second}.
As part of this approach, a pruning parameter $p$ is tuned to maximize the performance of the diarization, for which we obtained speaker labels for a small subset of our data\footnote{We thank the members of USC SCUBA lab for manually annotating speaker IDs and their corresponding speech segment timestamps}.

Once we obtained the speaker labels $S_1$ and $S_2$, the pitch of their corresponding segments was extracted and the ID \textit{Husband} was assigned to the label with lower median pitch and \textit{Wife} to the remaining label.
Manual inspection of a dozen sessions at random revealed acceptable accuracy of the ID assignment.
% \panos{some better quantification would be good...}
\vspace{-0.25cm}
% A sample session with speaker IDs, segments and time-stamps is shown in Fig.~\ref{fig:sample_interaction}.

% \begin{figure}[h]
% \centering
% \includegraphics[height=0.22\linewidth, width=0.9\linewidth]{sample_interaction.pdf}
% \caption{{Sample session of duration $T$ seconds with 6 speaker turns. $S_1$ and $S_2$ denote speaker IDs, $t_1, t_2, \ldots t_{16}$ denote speech segment time-stamps and $U_i$ denotes the $i^{th}$ utterance.}}
% \label{fig:sample_interaction}
% \end{figure}

\subsubsection{Automatic Speech Recognition}
We used the Kaldi \cite{povey2011kaldi} ASpIRE chain model\footnote{https://kaldi-asr.org/models/m1} but adapted the Language Model (LM) on related psychotherapy data in order to improve recognition accuracy.
Adaptation was performed by interpolating, with equal weights, 3 LMs that were trained on ASpIRE \cite{ko2017study}, cantab-TEDLIUM \cite{williams2015scaling} and a mix of Couples Therapy \cite{christensen2004} and Motivational Interviewing \cite{atkins2014scaling} corpora, using SRILM \cite{stolcke2002srilm}.
We built the session transcripts using 1-best hypotheses; spurious word insertions in noisy segments were eliminated with the help of word confusion heuristics such as number of confusions and confidence score.
\vspace{-0.25cm}

\subsection{Features}
\label{ssec:features}

\subsubsection{Acoustic Low-level Descriptors (A)}
\label{sssec:emobase}
Acoustic features have been shown to be useful in prior work as markers of  suicide risk \cite{venek2017adolescent, cummins2015review, scherer2013investigating} and depression \cite{cummins2015review, sanchez2011using}.
We used OpenSMILE \cite{eyben2013recent} for extracting the standard openEAR Emobase feature set \cite{eyben2009openear} from the speech segments of each speaker separately.
This set includes various prosody features (pitch, intensity \textit{etc}), voice quality features (jitter, shimmer \textit{etc}), and spectral features (MFCCs, line spectral frequencies \textit{etc}).
Then we took six statistical functionals (e.g. mean, standard deviation) over the low-level descriptors to obtain 228 session-level features for each speaker.
\vspace{-0.25cm}

\subsubsection{Acoustic Behavior Embeddings (E)}
\label{sssec:emobeh}
%\vspace{-0.125cm}
Speaker behaviors and mental states such as \textit{blame}, \textit{negativity} and \textit{depression} can provide clues about their suicidal risk \cite{proctor1994risk}.
To capture this information, we extract behavior embeddings from the acoustic channel using the reduced Context-Dependent model proposed in \cite{li2019linking} which employs emotions as primitives for facilitating behavior quantification and is trained on the Couples Therapy corpus \cite{christensen2004}.

We employ two sets of embeddings: (1) a 5-dimensional score vector $s$ from the model's final prediction layer corresponding to 5 behaviors: \textit{Acceptance, Blame, Positive, Negative, Sadness}, with higher score denoting stronger behavior, and (2) a 128-dimensional hidden representation $h$ from the model's penultimate layer.
We extract these from 26 different model configurations and concatenate them in the following fashion $[h_1, h_2, \ldots h_{26}, s_1, s_2, \ldots s_{26}]$, where $h_i$ and $s_i$ denote the hidden representation and score vector of the $i^{th}$ configuration.
This gives us 3458 session-level features per speaker.
% \vspace{-0.125cm}

\subsubsection{Lexical Cues (L)}
\label{sssec:lexical}
%\vspace{-0.25cm}
Since emotion expressed in language has been reported previously \cite{venek2017adolescent} to be associated with suicidal risk, we computed count-based statistics of LIWC \cite{pennebaker2001linguistic} \textit{positive} and \textit{negative} emotion words from the session transcripts.
6 lexical features were extracted: the proportions of both emotions in the speaker's language throughout the session, followed by the log-ratios of the speaker's and their partner's proportions for all 4 combinations of emotions (e.g. log-ratio of speaker-\textit{negative}-proportion to partner-\textit{positive}-proportion, etc.).
\vspace{-0.25cm}

\subsubsection{Turn-Taking Cues (T)}
\label{sssec:turn}
% %\vspace{-0.25cm}
The dynamics of turn-taking and pausing during an interaction have been linked to suicidal risk and psychological distress in multiple studies \cite{venek2017adolescent,cummins2015review,pestian2017machine}.
To capture them, we extract features relating to speech duration, number of words, speech rate and pause for every speaker and also compute differences between the speaker's and their partner's features during a turn change.
This is performed locally in every turn as well as globally over the entire session where applicable.
For each of the local features, we derive first and second-order temporal differences (delta ($\Delta$), delta-delta ($\Delta$$\Delta$)) and compute 9 session-level statistical functionals such as \textit{min}, \textit{median}, and quartiles on top of them.
This results in 167 session-level features.
\vspace{-0.25cm}

\section{Methodology}
\label{sec:methodology}
\vspace{-0.5cm}
\begin{table}[t]
\centering
\begin{tabular}{cc}
\toprule
\textbf{Correlation} & \textbf{Feature}               \\ \midrule
-0.219               & sadness score from $s_{18}$ (E)                        \\ 
-0.196               & sadness score from $s_{17}$ (E)                        \\ 
-0.175               & positive score from $s_{18}$ (E)                       \\ 
0.173                & std-dev($\Delta$ no. of words per turn) (T) \\ 
0.172                & std-dev(no. of words per turn) (T) \\ \bottomrule
\end{tabular}
\caption{Top 5 features most correlated with suicidal risk (feature set in parentheses). All correlations statistically significant $(p<0.05)$}
\label{tab:top5}
\end{table}

\begin{table}[t]
\centering
\begin{tabular}{ccc}
\toprule
\multicolumn{1}{c}{\textbf{Set}} & \multicolumn{1}{c}{\textbf{Correlation}} & \multicolumn{1}{c}{\textbf{Feature}} \\ \midrule
A & 0.1696 & line spectral pair coefficient \\
E & -0.219 & sadness score from $s_{18}$ \\
L & 0.105 & log($\frac{\text{speaker-\textit{negative}-proportion}}{\text{speaker-\textit{positive}-proportion}}$) \\
T & 0.173 & std-dev($\Delta$ no. of words per turn) \\ \bottomrule
\end{tabular}
\caption{Feature most correlated with suicidal risk, for every feature set. All correlations are statistically significant $(p<0.05)$}
\label{tab:topset}
\end{table}

\subsection{Analysis}
\label{ssec:analysis}
In order to understand how our features are related to suicidal risk, we compute their Spearman's rank correlation with the degree of risk.
Tables \ref{tab:top5} and \ref{tab:topset} show the top 5 most correlated features overall, and the most correlated feature per feature set respectively.
We see that behavior features have the highest correlation despite not being fine-tuned on our domain; this suggests the benefit of transfer learning from Couples Therapy.
Along with turn-taking cues, they appear to be the most promising features.
However, we observe weak monotonic relations in our features, which underscores the need for multimodal approaches, while suggesting possible non-linearities in how our features relate to suicidal risk.

\subsection{Experimental Setup}
\label{ssec:experiments}
While the primary focus of this work is to accurately classify the 3 degrees of a person's suicidal risk, we are also interested in investigating the 2 constituent \dq{one-versus-rest} binary classification scenarios.
These could be of importance in scenarios where the goal is to distinguish no risk from some risk or to identify and isolate \textit{attempt}, the most important risk factor as mentioned in Sec.~\ref{sec:data}.
Hence, we perform classification experiments for the following 3 scenarios:
\begin{enumerate}[itemsep=0em, leftmargin=*]
    \item Degree of Risk: \textit{none} vs \textit{ideation} vs \textit{attempt}
    \item No-Risk vs Risk: \textit{none} vs \{\textit{ideation} / \textit{attempt}\}
    \item Non-Severe vs Severe Risk: \{\textit{none} / \textit{ideation}\} vs \textit{attempt}
\end{enumerate}

Behavior expression patterns are known to vary across speakers of different genders \cite{heavey1993gender} and are also likely to be influenced by the nature and topic of the interaction \cite{baucom2010gender}.
\textit{RFL} sessions, for instance, are typically marked by long, introspective monologues whereas \textit{Conflict} sessions involve vigorous back-and-forth exchanges over the marital issue.
To examine whether these factors (gender, topic) have an impact on the classification performance, we employ the following data partitioning schemes in our experiments:
\begin{enumerate}[itemsep=0em, leftmargin=*]
    \item \textit{None}: Same model for all speakers, sessions (1 model)
    \item \textit{Gender}: Separate models for Husband, Wife (2 models)
    \item \textit{Content}: Separate models in \textit{RFL}, \textit{Conflict} (2 models)
    \item \textit{Demand}: Separate models for Husband, Wife in \textit{RFL}; separate models for Wife in \textit{W-Conflict}, \textit{H-Conflict}; same model for Husband in \textit{Conflict} (5 models)
\end{enumerate}
The \textit{Demand} partition is designed based on findings in \cite{baucom2010gender}.

\subsection{Classification}
\label{ssec:classification}
A Support Vector Machine (SVM) was used as the classifier and multimodal features were tested through feature-level fusion.
Feature dimensionality reduction was applied using Principal Component Analysis such that 95\% of the total energy was retained.
We applied sample weighting to address class imbalance and tuned hyperparameters such as feature normalization scheme (min-max, z-score), SVM kernel (linear, rbf), SVM penalty $C$ and rbf influence $\gamma$ (both $10^{-5}, 10^{-4.5}, \ldots 10^{5}$) to optimize the classifier.

We used leave-one-couple-out cross-validation where, in fold $i$, couple $C_i$ was picked as the test split and the remaining couples $C_j, j$$\neq$$i$ were randomly assigned to an $80$:$20$ train:validation split such that both splits had similar label distributions and no couple appeared in more than 1 split.
The classifier was then trained on the train split, optimized on the validation split and used to predict the suicidal risk of the speaker(s) in couple $C_i$.
Since our dataset contains 62 couples, this procedure was repeated for all 62 folds and results were accumulated from all the folds.

Performance was evaluated using macro-average recall of classification in order to account for class imbalance.
To determine whether there were statistically significant ($p < 0.05$) differences between our results and chance, we ran the McNemar's test for binary classifications and the Stuart-Maxwell test for the 3-class scenario.
In partition experiments, a separate classifier was created for each partition and at test time, the appropriate one was used.

\section{Results \& Discussion}
\label{sec:results}

\begin{table}[t]
\centering
\begin{tabular}{cccc}
\toprule
\textbf{Scenario} & \textbf{\begin{tabular}[c]{@{}c@{}}Degree\\ of Risk\end{tabular}} & \textbf{\begin{tabular}[c]{@{}c@{}}No-Risk\\ vs Risk\end{tabular}} &  \textbf{\begin{tabular}[c]{@{}c@{}}Non-Severe vs\\ Severe Risk\end{tabular}} \\ \midrule
Chance & 33.33 & 50.00 & 50.00 \\
Best system & 39.60$^*$ & 60.32$^*$ & 56.77$^*$ \\ \midrule
Feature Sets & A + E + T & E + L + T & L \\ \bottomrule
% Partition & {Gender} & {Demand} & {None} \\ 
\end{tabular}
\caption{Average Recall \% and features of best system in different classification scenarios. * denotes statistically significant ($p < 0.05$)}
\label{tab:results_task}
\end{table}

Table \ref{tab:results_task} shows the best classification performance and its corresponding features for each scenario.
We see that our system performs 13\% - 20\% (relative) better than chance in all 3 scenarios, with differences being statistically significant ($p$ $<$ $0.05$) and each feature set contributing to the best system in at least result.
In line with observations in Sec.~\ref{ssec:analysis}, the best features in 2 out of 3 scenarios consist of behavior and turn-taking cues. 
Our findings are similar to those of previous works \cite{venek2017adolescent,gideon2019emotion} which found affect-based and interaction dynamics cues to be important in identifying suicidal risk markers.
This suggests that emotional regulation can serve as a useful source of features for these tasks.

Separating no-risk from risk appears to be an easier problem than isolating severe risk from the rest.
One explanation for this could be that the communication patterns of \textit{ideation} are more similar to \textit{attempt} than \textit{none} and are, thus, harder to distinguish.
Another reason, however, could be the high class imbalance arising from combining \textit{none} and \textit{ideation} into a single class, leaving the \textit{attempt} class  with much fewer samples to train on.
Therefore, a more in-depth investigation on a larger dataset is required in order to better understand this issue.

Table \ref{tab:results_part} shows the best classification performance of each partition in every scenario.
We see that partition-based systems perform better than \textit{None} in 2 out of 3 scenarios, despite having lesser data to train on for each classifier.
Partitioning based on \textit{Gender} appears to be useful for risk classification, in general, whereas \textit{Demand} seems to be suited only to the scenario with severe risk; \textit{Content}, on the other hand, does not provide any noticeable benefits over other partitions.
This suggests that partitioning based on gender in conjunction with topic is a promising direction for further exploration, especially in mixture-of-experts frameworks.
\vspace{-0.125cm}

\begin{table}[t]
\centering
\begin{tabular}{cccc}
\toprule
\textbf{Partition} & \textbf{\begin{tabular}[c]{@{}c@{}}Degree\\ of Risk\end{tabular}} & \textbf{\begin{tabular}[c]{@{}c@{}}No-Risk\\ vs Risk\end{tabular}} & \textbf{\begin{tabular}[c]{@{}c@{}}Non-Severe vs\\ Severe Risk\end{tabular}} \\ \midrule
Chance & 33.33 & 50.00 & 50.00 \\ \midrule
\textit{None} & 37.78 & \textbf{60.32}$^*$ & 56.48$^*$ \\
\textit{Gender} & \textbf{39.60}$^*$ & 59.03 & 54.86$^*$ \\
\textit{Content} & 37.46 & 58.28 & 52.10$^*$ \\
\textit{Demand} & 36.60$^*$ & 53.90 & \textbf{56.77}$^*$ \\ \bottomrule
\end{tabular}
\caption{Best average recall \% of partitions in different classification scenarios. \textbf{Bold} value is best performance overall in that scenario. \hspace{1cm} * denotes statistically significant difference from chance ($p < 0.05$)}
\label{tab:results_part}
\end{table}

\section{Conclusions \& Future Work}
\label{sec:conclusion}
%%\vspace{-0.125cm}
In this work, we demonstrated the feasibility of an automated, multimodal approach to classifying military couples' suicidal risk by observing their conversations at home in real-world noisy conditions.
We also found that conditioning on speaker gender and discussion topic could benefit the classification of risk categories in specific scenarios.
For our future work, we will employ dynamic processing of cues instead of session-level aggregates and leverage information from the visual modality.
We also plan on incorporating frameworks that explicitly characterize couples dynamics, such as entrainment and influence, into our modeling.
% \vspace{-0.125cm}

\section{Acknowledgments}
\label{sec:ack}
The U.S. Army Medical Research Acquisition Activity is the awarding and administering acquisition office.
This work is supported by the Office of Assistant Secretary of Defense for Health Affairs through the Psychological Health and Traumatic Brain Injury Research Program under Award No. W81XWH-15-1-0632.
The funders had no role in study design, data collection and analysis, decision to publish, or preparation of the manuscript.

% \newpage

\normalsize
\bibliographystyle{IEEEbib}
\bibliography{strings,refs_orig}

\end{document}